\documentclass[twocolumn,showpacs,preprintnumbers,amsmath,amssymb,superscriptaddress]{revtex4-1}
\usepackage[maccyr]{inputenc}
\usepackage[T2A]{fontenc}
\usepackage[english,russian]{babel}
\usepackage{bm}% bold math  twocolumn,
\usepackage{epsfig,amssymb,amsmath,bm}

\begin{document}

\title{On Calculation of Amplitudes in Quantum Electrodynamics}
\author{K. S. Karplyuk}
\email{karpks@hotmail.com}
 \affiliation{Department of Radiophysics, Taras Shevchenko University, Academic
Glushkov prospect 2, building 5, Kyiv 03122, Ukraine}
\author{O. O. Zhmudskyy}\email{ozhmudsky@physics.ucf.edu}
 \affiliation{Department of Physics, University of Central Florida, 4000 Central Florida Blvd. Orlando, FL, 32816 Phone: (407)-823-4192}
\begin{abstract}

A new method of calculation of amplitudes of different processes in quantum electrodynamics is proposed.
The method does not use the Feynman technique of trace of product of matrices calculation. The
method strongly simplifies calculation of cross sections for different  processes. The
effectiveness of the method is shown on the cross-section calculation of Coulomb scattering, Compton scattering  and electron-positron annihilation.
\end{abstract}


\maketitle

The most labor-intensive part of calculation of cross sections for different processes in quantum electrodynamics is the amplitude calculation for these processes. Such calculations for non-polarized electrons, that is, electrons with no definite incoming or outgoing spin states, can be simplified by the Feynman trace technology of the traces of products of $\gamma$-matrices calculation  \cite{f}.  In this paper we propose a method which strongly simplifies the amplitude calculation for any elementary processes.

Let us start from the identity which is satisfied for matrices with arbitrary complex elements:
\begin{equation}
\chi M\psi=\mathrm{Sp}\,\hat{\psi}\hat{\chi} M.
\end{equation}
Here $M$ --- arbitrary square matrix, $\psi$ --- matrix-column, $\chi$ --- matrix-row. The square matrix $\hat{\psi}$  has only one nonzero column (let it be the n-th) which is equal to $\psi$. The square matrix $\hat{\chi}$  has only one n-th nonzero row which is equal to $\chi$.
In this paper we will use $\hat{\psi}$ matrix with the first non-zero column and the $\hat{\chi}$ matrix with the first non-zero row. According to (1)
\begin{equation}
\bar{u}^fMu^i=\mathrm{Sp}\,\hat{u}^i\hat{\bar{u}}^f M=\mathrm{Sp}\,\hat{u}^i\hat{u}^{\dag f}\gamma^0 M.
\end{equation}
Bispinors ${u}^i$ and ${u}^f$ represent the initial and final state of the fermions with momenta $\bm{p}^{i,f}$ and spins aligned along the unit vectors $\bm{s}^{i,f}$. They can be written as follows
\begin{equation}
u^{i,f}=\sqrt{\frac{p_0^{i,f}+mc}{2p_0^{i,f}V}}\!\left(1+\frac{\bm{p}^{i,f}\bm{\varsigma}_1}{p_0^{i,f}+mc}\right)\!\! \frac{1+i\bm{s}^{i,f}\bm{\varsigma}_2}{\sqrt{2(1+s_z^{i,f})}}
\left[\!\!\begin{array}{c}
1\\0\\
0\\0
\end{array}\!\!\right].
\end{equation}
These bispinors are normalized for one particle in a volume $V$, that is $\bar{u}^{i,f}\gamma^0u^{i,f}=u^{{i,f}\dag}
u^{i,f}=\displaystyle\frac{1}{V}$.

Here and below for brevity the following designations are used
\begin{gather*}
\bm{c}\bm{\gamma}=c_x\gamma^1+c_y\gamma^2+c_z\gamma^3, 
\end{gather*}
\begin{gather*}
\bm{c}\bm{\varsigma}_1=c_x\gamma^0\gamma^1+c_y\gamma^0\gamma^2+c_z\gamma^0\gamma^3,\\
\bm{c}\bm{\varsigma}_2=c_x\gamma^2\gamma^3+c_y\gamma^3\gamma^1+c_z\gamma^1\gamma^2,\\
\bm{c}\bm{\pi}=c_x\gamma^0\gamma^2\gamma^3+c_y\gamma^0\gamma^3\gamma^1+c_z\gamma^0\gamma^1\gamma^2,\\
\pi^0=\gamma^1\gamma^2\gamma^3,\hspace{7mm}\hat{\iota}=\gamma^0\gamma^1\gamma^2\gamma^3,
\end{gather*}
where matrices $\gamma$ are used in standard Dirac-Pauli representation. Let us use (3), and calculate the product $\hat{u}^i\hat{\bar{u}}^f$:
\begin{gather}
\hat{u}^i\hat{\bar{u}}^f=\frac{1}{4V}\sqrt{\frac{(p_0^i+mc)(p_0^f+mc)}{p_0^ip_0^f(1+s_z^i)(1+s_z^f)}}
\times\nonumber\\
\times\!{\Bigl(1\!+\!\frac{\bm{p}^i\bm{\varsigma}_1}{p_0^i+mc}\Bigr)\!\!
\Bigl[a_0(1\!+\!\gamma^0)\!+\!\bm{a}(\bm{\pi}\!+\!\bm{\varsigma}_2)\Bigr]\!\!
\Bigl(1\!+\!\frac{\bm{p}^f\bm{\varsigma}_1}{p_0^f+mc}\Bigr)\gamma_0}\!=\nonumber\\
=\frac{1}{4V}\sqrt{\frac{(p_0^i+mc)(p_0^f+mc)}{p_0^ip_0^f(1+s_z^i)(1+s_z^f)}}
\Bigr\{a_0(1+\gamma^0)+\bm{a}(\bm{\pi}+\bm{\varsigma}_2)+\nonumber\\
+\frac{(\bm{a}\cdot\bm{p}^i)(\hat{\iota}-\pi^0)+
(a_0\bm{p}^i-\bm{a}\times\bm{p}^i)
(\bm{\varsigma}_1-\bm{\gamma})}{p_0^i+mc}-\Bigl.\nonumber\\
-\frac{(\bm{a}\cdot\bm{p}^f)(\hat{\iota}+\pi^0)+
(a_0\bm{p}^f+\bm{a}\times\bm{p}^f)
(\bm{\varsigma}_1+\bm{\gamma})}{p_0^f+mc}-\nonumber\\
-\frac{a_0(\bm{p}^i\cdot\bm{p}^f)-\bm{a}\cdot
(\bm{p}^i\times\bm{p}^f)}{(p_0^i+mc)(p_0^f+mc)}(1-\gamma^0)-\nonumber\\
\Bigr.-\!\frac{a_0(\bm{p}^i\!\times\!\bm{p}^f)\!+\!
(\bm{p}^i\!\cdot\!\bm{p}^f)\bm{a}\!-\!(\bm{a}\cdot\bm{p}^f)\bm{p}^i\!-\!
(\bm{a}\!\cdot\!\bm{p}^i)\bm{p}^f}
{(p_0^i+mc)(p_0^f+mc)}(\bm{\pi}\!-\!\bm{\varsigma}_2)\!\Bigl\}.
\end{gather}
Here
\begin{gather}
a_0=\frac{i}{4}\bm{e}_z\!\cdot\!(\bm{s}^f\!\times\!\bm{s}^i)+\frac{1}{4}(1+\bm{s}^i\!\cdot\!\bm{s}^f+\bm{e}_z\!\cdot\!\bm{s}^i+\bm{e}_z\!\cdot\!\bm{s}^f),\\
\bm{a}\!=\!\frac{i}{4}\bigl[\bm{e}_z\!\!+\!\bm{s}^i\!\!+\!\bm{s}^f\!\!+\!(\bm{e}_z\!\cdot\!\bm{s}^f)\bm{s}^i\!\!+\!(\bm{e}_z\!\cdot\!\bm{s}^i)\bm{s}^f\!-
\!\bm{e}_z(\bm{s}^i\!\cdot\!\bm{s}^f)\bigr]\!+\nonumber\\
+\frac{1}{4}\bigl(\bm{s}^f\!\!\times\bm{s}^i-\bm{e}_z\times\bm{s}^f+\bm{e}_z\times\bm{s}^i\bigr),
\end{gather}
where $\bm{e}_z$ is the unit vector along the $z$ axis. Note that $a_0$ and $\bm{a}$ depend on the direction of the spins $\bm{s}^i$ and $\bm{s}^f$ only, and do not depend on energies and momenta of initial and final fermions.

As far as $\hat{u}^i\hat{\bar{u}}^f$is known, the trace of the matrix $\hat{u}^i\hat{\bar{u}}^fM$ can be calculated.
In the general case matrix $M$ has a form:
\begin{equation}
M=I+V_0\gamma^0+\bm{V}\bm{\gamma}+W_0\pi^0+\bm{W}\bm{\pi}+
\bm{E}\bm{\varsigma}_1+\bm{B}\bm{\varsigma}_2+J\hat{\iota}.
\end{equation}
The unit matrix is the only one of the sixteen Dirac matrices which has nonzero trace. That is why in the matrix product of $\hat{u}^i\hat{\bar{u}}^f$ and $M$ it is enough to take into account only those terms which are proportional to the unit matrix. Such terms appear only for a multiplication of the same matrices.  This remark simplifies the multiplication and trace calculation:
\begin{gather}\bar{u}^fMu^i=\mathrm{Sp}\,\hat{u}^i\hat{\bar{u}}^fM=\nonumber\\
=\sqrt{\frac{(p_0^i+mc)(p_0^f+mc)}{p_0^ip_0^f(1+s_z^i)(1+s_z^f)}}
\frac{1}{V}\Bigl(a_0K_0-\bm{a}\cdot\bm{K}\Bigr).
\end{gather}
In three dimensions the scalar $K_0$ and the pseudo-vector $\bm{K}$ can be written as
\begin{gather}
K_0=(I+V_0)+\frac{\bm{p}^i\cdot(\bm{E}+\bm{V})}{p_0^i+mc}
-\frac{\bm{p}^f\cdot(\bm{E}-\bm{V})}{p_0^f+mc}+\nonumber\\
+\frac{(\bm{p}^i\times\bm{p}^f)
\cdot(\bm{W}-\bm{B})-(\bm{p}^i\cdot\bm{p}^f)(I-V_0)}{(p_0^i+mc)(p_0^f+mc)},\\
\bm{K}\!=\!(\bm{W}\!+\!\bm{B})+\frac{\bm{p}^i(W_0\!+\!J)+\bm{p}^i\!\times\!(\bm{E}\!+\!\bm{V})}{p_0^i+mc}+\nonumber\\
+\frac{\bm{p}^f(W_0\!-\!J)+ \bm{p}^f\!\times\!(\bm{E}\!-\!\bm{V})}{p_0^f+mc}+\nonumber\\
+\frac{\bm{p}^i[\bm{p}^f\!\cdot\!(\bm{W}\!\!-\!\!\bm{B})]\!+\!\bm{p}^f[\bm{p}^i\!\cdot\!(\bm{W}\!\!-\!\!\bm{B})]\!-\!
(\bm{p}^i\!\cdot\!\bm{p}^f)(\bm{W}\!\!-\!\!\bm{B})}{(p_0^i+mc)(p_0^f+mc)}-\nonumber\\
-\frac{(\bm{p}^i\times\bm{p}^f)(I-V_0)}{(p_0^i+mc)(p_0^f+mc)}.
\end{gather}
Note that $K_0$ and $\bm{K}$ depend on energies and momenta of initial and final fermion states only and do not depend on their polarization  $\bm{s}^i$ and $\bm{s}^f$.

In order to evaluate the probability and cross-section of the process the square of the amplitude of transaction must be calculated.
Let us calculate $|(a_0K_0-\bm{a}\bm{K})|^2$ supposing that all coefficients in (7) are real. This is usually the case in quantum electrodynamics.
Thus the modulus square $|\bar{u}^fMu^i|^2$ is
\begin{gather}
|\bar{u}^fMu^i|^2=\nonumber\\
=\!{\frac{(p_0^i+mc)(p_0^f+mc)}{p_0^ip_0^f(1\!+s_z^i)(1\!+s_z^f)V^2}}(a_0K_0\!-\!\bm{a}\!\cdot\!\bm{K})(a_0^*K_0\!-\!\bm{a}^*\!\cdot\!\bm{K})\!=\nonumber\\
=\!{\frac{(p_0^i\!+\!mc)(p_0^f\!+\!mc)}{8V^2p_0^ip_0^f}}\Bigl[(1\!+\!\bm{s}^i\bm{s}^f)K_0^2\!+\!(1\!-\!\bm{s}^i\bm{s}^f)\bm{K}\!\cdot\!\bm{K}\!+\!\nonumber\\
+2(\bm{s}^i\bm{K})(\bm{s}^f\bm{K})+2(\bm{s}^i\times\bm{s}^f)\bm{K}K_0\Bigr].
\end{gather}
Expressions (8)-(11) are universal. They determine $\bar{u}^fMu^i$ and $|\bar{u}^fMu^i|^2$ for any processes in the quantum electrodynamics. Different processes differ by matrix $M$ only. The only thing we need to do in order to calculate $\bar{u}^fMu^i$ and $|\bar{u}^fMu^i|^2$ is to represent the interaction matrix $M$ in a form (7). Then substitute coefficients from matrix (7) into expressions (9)-(10). These expressions entirely determine $K_0$ and $\bm{K}$. Equations (8) and (11) give the algebraic expressions for $\bar{u}^fMu^i$ and $|\bar{u}^fMu^i|^2$. All that remains is the simplification of $\bar{u}^fMu^i$ and $|\bar{u}^fMu^i|^2$ as much as possible.

Expression (11) determines the square of the amplitude of the process which corresponds to any desirable spin states of the incoming and outgoing fermions. This expression explicitly represents dependence on the fermion polarization that is why this dependence can be easily analyzed. If a detector is blind to polarization, i.e. for the final state both polarizations of fermions in $\bm{s}^f$ direction and in $-\bm{s}^f$ direction are registered, expression (11) must include the sum for both directions $\bm{s}^f$ and $-\bm{s}^f$:
\begin{equation}
|\bar{u}^fMu^i|^2=\frac{(p_0^i+mc)(p_0^f+mc)}{4V^2p_0^ip_0^f}[K_0^2+\bm{K}\cdot\bm{K}].
\end{equation}
Let us demonstrate the effectiveness of the above method on three examples:  Coulomb scattering, Compton scattering and electron - positron annihilation.

\subsection{Coulomb Scattering}
We will define the cross section of an electron of charge $e$ scattering on the Coulomb
center of charge $Ze$ versus the square of the amplitude $|\bar{u}^f\gamma^0u^i|^2$ in a usual way:
\begin{gather}
\frac{d\sigma}{d\Omega}=\frac{(2Zr_0mc^2)^2}{(2p\sin\frac{\theta}{2})^4}\left(\frac{Vp_o}{c}\right)^2|\bar{u}^f\gamma^0u^i|^2=\nonumber\\
=\left(\frac{Zr_0}{2}\right)^2\left(\frac{c}{v}
\frac{1}{\sin\frac{\theta}{2}}\right)^4\left(1-\frac{v^2}{c^2}\right)V^2|\bar{u}^f\gamma^0u^i|^2.
\end{gather}
Here $r_0$ is the classical electron radius, $\theta$ is the scattering angle. The amplitude square $|\bar{u}^f\gamma^0u^i|^2$ is defined by the universal expression (11), in which $K_0$ and $\bm{K}$ must be calculated for the matrix $M=\gamma^0$. Hence, in expressions (9)-(10) we must set  $V_0=1$. All other coefficients must be set  to zero. We must also take into account that $p_0^i=p_0^f=p_0$ because for the Coulomb  scattering  energy is conserved:
\begin{equation}
K_0=1+\frac{\bm{p}^i\cdot\bm{p}^f}{(p_0\!+\!mc)^2},\hspace{7mm}
\bm{K}=\frac{\bm{p}^i\times\bm{p}^f}{(p_0\!+\!mc)^2}.
\end{equation}
So, the cross section becomes:
\begin{gather}
\frac{d\sigma}{d\Omega}=\frac{1}{2}\left(\frac{Z r_0}{2}\right)^2\left(\frac{c}{v}
\frac{1}{\sin\frac{\theta}{2}}\right)^4\left(1-\frac{v^2}{c^2}\right)
\left(\frac{p_0+mc}{2p_0}\right)^2
\times\nonumber\\
\times\Bigl[(1\!+\!\bm{s}^i\!\cdot\!\bm{s}^f)K_0^2\!+\!(1-\bm{s}^i\!\cdot\!\bm{s}^f)K^2\!+\!
2(\bm{s}^i\!\cdot\!\bm{K})(\bm{s}^f\!\cdot\!\bm{K})\!+\nonumber\\
+2(\bm{s}^f\times\bm{s}^i)\cdot\bm{K}K_0\Bigr].
\end{gather}
In expression (15) $K_0$ and $\bm{K}$ are determined according to equation (14).

Expression (15) determines the differential cross section $d\sigma/d\Omega$ in the case with definite incoming and outgoing electron spin states. If the polarization of the final electron is not registered, expression (12) must be used:
\begin{equation}
\frac{d\sigma}{d\Omega}=\left(\frac{Z r_0}{2}\right)^{\!\!2}\!\!\!\left(\frac{c}{v}
\frac{1}{\sin\frac{\theta}{2}}\right)^{\!\!4}\!\!\!\left(1\!-\!\frac{v^2}{c^2}\right)\!\!\!
\left(\frac{p_0\!+\!mc}{2p_0}\right)^{\!\!2}\!\!\!(K_0^2+K^2).
\end{equation}
Expression $\displaystyle\left(\frac{p_0+mc}{2p_0}\right)^2 (K_0^2+\bm{K}\cdot\bm{K})$ can be simplified:
\begin{gather}
\left(\frac{p_0+mc}{2p_0}\right)^2(K_0^2+\bm{K}\cdot\bm{K})=\nonumber\\
=\left(\frac{p_0+mc}{2p_0}\right)^2 \left\{
\left[1+\frac{p^2\cos\theta}{(p_0+mc)^2}\right]^2+
\frac{p^4\sin^2\theta}{(p_0+mc)^4}\right\}=\nonumber\\
=\frac{p_0^2+p_0^2-p^2(1-\cos\theta)}{2p_0^4}=
\left(1-\frac{v^2}{c^2}\sin^2\frac{\theta}{2}\right)
\end{gather}
Recall that $\theta$ is an angle between $\bm{p}^i$ and $\bm{p}^f$.
After substituting (17) into (16) we come up with unpolarized cross section for Coulomb scattering:
\begin{equation}
\frac{d\sigma}{d\Omega}=\left(\frac{Z r_0}{2}\right)^2\left(\frac{c}{v}
\frac{1}{\sin\frac{\theta}{2}}\right)^4\left(1-\frac{v^2}{c^2}\right)
\left(1-\frac{v^2}{c^2}\sin^2\frac{\theta}{2}\right).
\end{equation}
This is the well-known Mott scattering cross section \cite{m}. Note that for the polarized cross section calculation (15) and the unpolarized cross section calculation (18) instead of the Feynman technique (of  trace of product  of matrices calculation) we use expressions  (9)-(12) which strongly simplify calculations.

\subsection{Compton Scattering}
It is well-known that Compton scattering in the first order of probability theory is represented by two Feynman diagrams. Call them $a$
and $b$. According to the $a$ diagram an electron absorbs a photon of frequency $\omega_1$ first, and then emits a photon of frequency $\omega_2$. According to the $b$ diagram an electron emits a  photon of frequency $\omega_2$ first, and then absorbs a photon of frequency $\omega_1$. The amplitudes for the two diagrams must be added and their sum squared.

Assume that the incoming electron is at rest, hence $\bm{p}^i=0$, $p_0^i=mc$. Let us express the scattering
cross section versus the square of the sum of the amplitudes $|\bar{u}^fMu^i|^2$ in a usual way:
\begin{gather}
\frac{d\sigma}{d\Omega}=r_0^2mcp_0^fV^2\left(\frac{\omega_2}{\omega_1}\right)^2|\bar{u}^fMu^i|^2.
\end{gather}
Here $r_0$ is the classical electron radius. Matrix $M$ for the two diagrams is
\begin{gather*}
M=\\
=\bm{\mathfrak{e}}_2\bm{\gamma}\frac{p_{0a}\gamma^0\!-\!\bm{p}_a\bm{\gamma}\!+\!mc}
{2mc\hbar k_1}\bm{\mathfrak{e}}_1\bm{\gamma}+\bm{\mathfrak{e}}_1\bm{\gamma}\frac{p_{0b}\gamma^0\!-\!\bm{p}_b\bm{\gamma}\!+\!mc}
{(-2mc\hbar k_2)}\bm{\mathfrak{e}}_2\bm{\gamma}.
\end{gather*}
Here $p_{0a}=mc+\hbar k_1$, $\bm{p}_a=\hbar\bm{k}_1$, $p_{0b}=mc-\hbar k_2$, $\bm{p}_b=-\hbar\bm{k}_2$, $\bm{k}_1$ and $\bm{k}_2$ are the wave vectors of photons $1$ and $2$, $k_1=\omega_1/c$, $k_2=\omega_2/c$, $\bm{\mathfrak{e}}_1$ and $\bm{\mathfrak{e}}_2$ are the unit vectors of polarization of   photons $1$ and $2$,
$\bm{\mathfrak{e}}_1\cdot\bm{k}_1=0$, $\bm{\mathfrak{e}}_2\cdot\bm{k}_2=0$. The polar angle in $d\Omega$ is measured from the $\bm{k}_1$ direction. Matrix multiplication in the expression for $M$ leads to the coefficients in equation (7):
\begin{gather*}
I=\frac{(\bm{\mathfrak{e}}_1\cdot\bm{\mathfrak{e}}_2)}{2\hbar}\Bigl(\frac{1}{k_2}-\frac{1}{k_1}\Bigr),\hspace{3mm}
V_{0}=\frac{(\bm{\mathfrak{e}}_1\cdot\bm{\mathfrak{e}}_2)}{2mc\hbar}\Bigl(\frac{p_{0a}}{ k_1}-\frac{p_{0b}}{ k_2}\Bigr),\nonumber\\
\bm{V}=\frac{(\bm{p}_a\cdot\bm{\mathfrak{e}}_1)\bm{\mathfrak{e}}_2-
\bm{\mathfrak{e}}_2\times(\bm{p}_a\times\bm{\mathfrak{e}}_1)}{2mc\hbar k_1}+\nonumber\\
+\frac{(\bm{p}_b\cdot\bm{\mathfrak{e}}_2)\bm{\mathfrak{e}}_1-
\bm{\mathfrak{e}}_1\times(\bm{p}_b\times\bm{\mathfrak{e}}_2)}{(-2mc\hbar k_2)},\nonumber\\
\bm{W}\!=\!\frac{(\bm{\mathfrak{e}}_1\times\bm{\mathfrak{e}}_2)}{2mc\hbar}\Bigl(\frac{ p_{0a}}{ k_1}+\frac{ p_{0b}}{ k_2}\Bigr),
\bm{B}=-\frac{(\bm{\mathfrak{e}}_1\times\bm{\mathfrak{e}}_2)}{2\hbar}\Bigl(\frac{1}{k_1}+\frac{1}{k_2}\Bigr),\nonumber\\
W_{0}=\frac{-\bm{\mathfrak{e}}_2\cdot(\bm{p}_a\times\bm{\mathfrak{e}}_1)}{2mc\hbar k_1}+\frac{\bm{\mathfrak{e}}_1\cdot(\bm{p}_b\times\bm{\mathfrak{e}}_2)}{2mc\hbar k_2},\hspace{3mm}\bm{E}=0,\hspace{3mm}J=0.
\end{gather*}
Substitution of these coefficients into (9)-(10) gives us expressions for $K_0$ and $\bm{K}$:
\begin{gather}
K_0=(I+V_0)+\frac{\bm{p}^f\cdot\bm{V}}{p_0^f+mc}=2\frac{\bm{\mathfrak{e}}_1\cdot\bm{\mathfrak{e}}_2}{2mc}-\nonumber\\
-\frac{1}{2mc}
\frac{\bm{p}^f\cdot[\bm{\mathfrak{e}}_2\times(\bm{\mathfrak{k}}_1\times\bm{\mathfrak{e}}_1)]}
{p_0^f+mc}-\frac{1}{2mc}
\frac{\bm{p}^f\cdot[\bm{\mathfrak{e}}_1\times(\bm{\mathfrak{k}}_2\times\bm{\mathfrak{e}}_2)]}{p_0^f+mc},\\
\bm{K}=(\bm{W}+\bm{B})+\frac{\bm{p}^fW_0-\bm{p}^f\times
\bm{V}}{p_0^f+mc}=\nonumber\\
=\frac{1}{2mc}
\frac{\bm{p}^f\times[\bm{\mathfrak{e}}_2\times(\bm{\mathfrak{k}}_1\times\bm{\mathfrak{e}}_1)]-
\bm{p}^f[\bm{\mathfrak{e}}_2\cdot(\bm{\mathfrak{k}}_1\times\bm{\mathfrak{e}}_1)]}{p_0^f+mc}+\nonumber\\
+\frac{1}{2mc}\frac{\bm{p}^f\times[\bm{\mathfrak{e}}_1\times(\bm{\mathfrak{k}}_2\times\bm{\mathfrak{e}}_2)]-
\bm{p}^f[\bm{\mathfrak{e}}_1\cdot(\bm{\mathfrak{k}}_2\times\bm{\mathfrak{e}}_2)]} {p_0^f+mc}.
\end{gather}
In expressions (20) and (21) $\displaystyle\bm{\mathfrak{k}}_1=\frac{\bm{k}_1}{k_1}$,
$\displaystyle\bm{\mathfrak{k}}_2=\frac{\bm{k}_2}{k_2}$. Expressions for $K_0$ and $\bm{K}$ together with (11) and (19) determine the polarized cross section. In order to get the unpolarized cross section, expressions (12) and (19) must be used. The sum $K_0^2+K^2$ can be simplified and expressed as
\begin{equation}
K_0^2+K^2=\frac{1}{2mc}\frac{1}{p_0^f+mc}\Bigl[4(\bm{\mathfrak{e}}_1\cdot\bm{\mathfrak{e}}_2)^2+\frac{(\omega_1-\omega_2)^2}{\omega_1\omega_2}\Bigr].
\end{equation}
Thus, the unpolarized cross section is
\begin{equation}
\frac{d\sigma}{d\Omega}=\frac{r_0^2}{4}\left(\frac{\omega_2}{\omega_1}\right)^2
\Bigl[4(\bm{\mathfrak{e}}_1\cdot\bm{\mathfrak{e}}_2)^2+\frac{(\omega_1-\omega_2)^2}{\omega_1\omega_2}\Bigr].
\end{equation}
This is the well-known Klein-Nishina scattering cross section \cite{kn}. As in the previous case we don't use the Feynman technique of trace of product of matrices calculation. Using the universal expressions (11) and (12) instead strongly simplified calculations.

\subsection{Annihilation}
Two Feynman diagrams represent the annihilation process in the first order of perturbation theory.
The first diagram, call it  \lq\lq a\rq\rq, corresponds to the process in which an incoming electron emits a photon $\gamma_1$ of frequency $\omega_1$, then a photon $\gamma_2$ of frequency $\omega_2$, and transfers to the state with negative energy.  The second diagram, call it \lq\lq b\rq\rq, corresponds to the process in which $\gamma_1$ and $\gamma_2$ interchange. In order to calculate the annihilation probability we must add the amplitudes of these processes and then square it.

For simplicity, assume that the electron is at rest, so $\bm{p}^i=0$, $p_0^i=mc$. Following Feynman, we treat the positron as an electron with negative energy moving backward in time. This electron has linear momentum and spin opposite in direction to the positron's momentum and spin. It allows us to describe the positron by the same bispinor (3) if we set up $p_0^f=-p^+$, $\bm{p}^f=-\bm{p}^+$, $\bm{s}^f=-\bm{s}^+$, where the index \lq\lq +\rq\rq\ designates positron quantities. The same substitutions have to be done in expressions (9) and (10). Let us express the annihilation cross section
versus the square of the sum of the amplitudes $|\bar{u}^fMu^i|^2$ in a usual way:
\begin{equation}
\frac{d\sigma}{d\Omega}=r_0^2\frac{m\hbar^2\omega_1^2p_0^+V^2}{|p^+|(p_0^++mc)c}|\bar{u}^fMu^i|^2.
\end{equation}
Matrix $M$, which corresponds to the sum of two diagrams is:
\begin{gather*}
M=\\
=\bm{\mathfrak{e}}_2\bm{\gamma}\frac{p_{0a}\gamma^0\!-\!\bm{p}_a\bm{\gamma}\!+\!mc}
{(-2mc\hbar k_1)}\bm{\mathfrak{e}}_1\bm{\gamma}+\bm{\mathfrak{e}}_1\bm{\gamma}\frac{p_{0b}\gamma^0\!-\!\bm{p}_b\bm{\gamma}\!+\!mc}
{(-2mc\hbar k_2)}\bm{\mathfrak{e}}_2\bm{\gamma}.
\end{gather*}
Here $p_{0a}=mc-\hbar k_1$, $\bm{p}_a=-\hbar\bm{k}_1$, $p_{0b}=mc-\hbar k_2$, $\bm{p}_b=-\hbar\bm{k}_2$, $\bm{k}_1$ and $\bm{k}_2$ are the wave vectors of the photons $1$ and $2$, $k_1=\omega_1/c$, $k_2=\omega_2/c$, $\bm{\mathfrak{e}}_1$  and $\bm{\mathfrak{e}}_2$ are the unit vectors of the polarization of the photons  $1$ and $2$,  $\bm{\mathfrak{e}}_1\cdot\bm{k}_1=0$, and $\bm{\mathfrak{e}}_2\cdot\bm{k}_2=0$. The polar angle in the $d\Omega$ is measured from the $\bm{k}_1$ direction.
After transformation of the matrix $M$ to the form (7) we can find coefficients in (7) in the reference frame in which incoming electron is at rest  $\bm{p}^i=0$:
\begin{gather*}
I+V_0=2\frac{\bm{\mathfrak{e}}_1\cdot\bm{\mathfrak{e}}_2}{2mc},\hspace{4mm} \bm{W}+\bm{B}=0,\hspace{4mm}\bm{E}=0,\hspace{4mm}J=0,\\
\bm{V}=-\frac{\bm{\mathfrak{e}}_1\times(\bm{\mathfrak{k}}_2\times\bm{\mathfrak{e}}_2)+
\bm{\mathfrak{e}}_2\times(\bm{\mathfrak{k}}_1\times\bm{\mathfrak{e}}_1)}{2mc}=\\
=-\frac{(\bm{\mathfrak{k}}_1+\bm{\mathfrak{k}}_2)(\bm{\mathfrak{e}}_1\cdot\bm{\mathfrak{e}}_2)-
\bm{\mathfrak{e}}_1(\bm{\mathfrak{e}}_2\cdot\bm{\mathfrak{k}}_1)
-\bm{\mathfrak{e}}_2(\bm{\mathfrak{e}}_1\cdot\bm{\mathfrak{k}}_2)}{2mc},\\
W_0=-\frac{\bm{\mathfrak{e}}_1\cdot(\bm{\mathfrak{k}}_2\times\bm{\mathfrak{e}}_2)
+\bm{\mathfrak{e}}_2\cdot(\bm{\mathfrak{k}}_1\times\bm{\mathfrak{e}}_1)}{2mc}.
\end{gather*}
Here  $\displaystyle\bm{\mathfrak{k}}_{1,2}=\frac{\bm{k}_{1,2}}{k_{1,2}}$, and $\displaystyle k_{1,2}=\frac{\omega_{1,2}}{c}$.
Using (9) and (10) scalar $K_0$ and pseudo-vector $\bm{K}$ can be calculated.
\begin{gather}
K_0=\frac{\bigl[2(p_0^f+mc)+\hbar(k_1\!+\!k_2)(1+\cos\theta)\bigr](\bm{\mathfrak{e}}_1\!\cdot\!\bm{\mathfrak{e}}_2)}{2mc(p_0^f+mc)}-\nonumber\\
-\frac{\hbar(k_1\!+\!k_2)(\bm{\mathfrak{k}}_1\!\cdot\!\bm{\mathfrak{e}}_2)(\bm{\mathfrak{k}}_2\!\cdot\!\bm{\mathfrak{e}}_1)}{2mc(p_0^f+mc)}.\\
\bm{K}=\frac{\bm{p}^f\!\times\bigl[\bm{\mathfrak{e}}_1\times(\bm{\mathfrak{k}}_2\!\times\!\bm{\mathfrak{e}}_2)\bigr]+
\bm{p}^f\!\times\bigl[\bm{\mathfrak{e}}_2\times(\bm{\mathfrak{k}}_1\!\times\!\bm{\mathfrak{e}}_1)\bigr]}{2mc(p_0^f+mc)}+\nonumber\\
+\frac{\bm{p}^f\bigl[\bm{\mathfrak{e}}_1\cdot(\bm{\mathfrak{e}}_2\!\times\!\bm{\mathfrak{k}}_2)\bigr]+
\bm{p}^f\bigl[\bm{\mathfrak{e}}_2\cdot(\bm{\mathfrak{e}}_1\!\times\!\bm{\mathfrak{k}}_1)\bigr]}{2mc(p_0^f+mc)}.
\end{gather}
These expression for  $K_0$ and $\bm{K}$ together with (11) and (24) determine the polarized cross section (both the electron and the positron have a given direction of spin). In order to calculate the unpolarized cross section expressions (12) and (24) must be used.
The sum $K_0^2+K^2$ can be reduced to
\begin{equation}
K_0^2+K^2=\frac{1}{2mc}\frac{1}{p_0^f+mc}\Bigl[4(\bm{\mathfrak{e}}_1\cdot\bm{\mathfrak{e}}_2)^2-\frac{(\omega_1+\omega_2)^2}{\omega_1\omega_2}\Bigr].
\end{equation}
Thus the unpolarized cross section is:
\begin{equation}
\frac{d\sigma}{d\Omega}=\frac{r_0^2}{4}\frac{\hbar^2k_1^2}{|p^+|(p_0^++mc)}
\Bigl[\frac{(\omega_1+\omega_2)^2}{\omega_1\omega_2}-4(\bm{\mathfrak{e}}_1\cdot\bm{\mathfrak{e}}_2)^2\Bigr].
\end{equation}
This result entirely coincides with the one calculated by the technique of trace of product of matrices calculation \cite{f} and with the one calculated by Dirac \cite{d}.

As it was shown in the above examples, the method proposed in this paper allows us to strongly simplify
calculation of the polarized and unpolarized fermion cross sections in quantum electrodynamics.
It is free from the necessity of calculation of trace of product of great amount of matrices.


\begin{thebibliography}{12}
\expandafter\ifx\csname natexlab\endcsname\relax\def\natexlab#1{#1}\fi
\expandafter\ifx\csname bibnamefont\endcsname\relax
  \def\bibnamefont#1{#1}\fi
\expandafter\ifx\csname bibfnamefont\endcsname\relax
  \def\bibfnamefont#1{#1}\fi
\expandafter\ifx\csname citenamefont\endcsname\relax
  \def\citenamefont#1{#1}\fi
\expandafter\ifx\csname url\endcsname\relax
  \def\url#1{\texttt{#1}}\fi
\expandafter\ifx\csname urlprefix\endcsname\relax\def\urlprefix{URL }\fi
\providecommand{\bibinfo}[2]{#2}
\providecommand{\eprint}[2][]{\url{#2}}


\bibitem[{\citenamefont{{Feynman}}(1961)}]{f}
\bibinfo{author}{\bibfnamefont{R.~P.} \bibnamefont{{Feynman}}},
\emph{\bibinfo{title}{Quantum electrodynamics}} (\bibinfo{publisher}{Benjamin, New York},
\bibinfo{year}{1961}).

\bibitem[{\citenamefont{{Mott}}(1929)}]{m}
\bibinfo{author}{\bibfnamefont{N.~F.} \bibnamefont{{Mott}}},
  \bibinfo{journal}{Proc. Roy. Soc.} \textbf{\bibinfo{volume}{A124}},
  \bibinfo{pages}{425} (\bibinfo{year}{1929}).

\bibitem[{\citenamefont{{Klein} and {Nishina}}(1929)}]{kn}
\bibinfo{author}{\bibfnamefont{O.} \bibnamefont{{Klein}}}
  \bibnamefont{and} \bibinfo{author}{\bibfnamefont{Y.}
  \bibnamefont{{Nishina}}},
  \bibinfo{journal}{Z. f. Phys.} \textbf{\bibinfo{volume}{52}},
  \bibinfo{pages}{853} (\bibinfo{year}{1929}).

\bibitem[{\citenamefont{{Dirac}}(1930)}]{d}
\bibinfo{author}{\bibfnamefont{P.~A.~M.} \bibnamefont{{Dirac}}},
  \bibinfo{journal}{Proc. Cambridge Phil. Soc.} \textbf{\bibinfo{volume}{26}},
  \bibinfo{pages}{361} (\bibinfo{year}{1930}).

\end{thebibliography}
\end{document}